\documentclass{elsart}
\usepackage{natbib}
\usepackage{graphicx}% Include figure files

\begin{document}
\runauthor{Kimura and Bonasera}
\begin{frontmatter}
%\title{Controlling fusion at astrophysical energies \\through the electrons chaotic motion}
\title{Influence of the Electronic Chaotic Motion \\on the Fusion Dynamics \\ at Astrophysical Energies} 
\author[infn-lns]{Sachie Kimura}
and 
%\author[infn-lns]{Sachie Kimura\thanksref{Someone}}
\author[infn-lns]{Aldo Bonasera}
%\thanks[X]{This is the history of the paper, etc etc}

\address[infn-lns]{Laboratorio Nazionale del Sud, INFN,
via Santa Sofia, 62, 95123 Catania, Italy}

\begin{abstract}
We perform semi-classical molecular dynamics simulations 
%to assess 
of screening by bound electrons in low energy nuclear 
reactions. In our simulations quantum effects corresponding 
to the Pauli and Heisenberg principle are enforced by constraints. 
In addition to the well known adiabatic and sudden limits,  
we propose a new "dissipative limit" which is expected 
to be important not only at high energies but in the extremely low energy region. 
The dissipative limit is associated with the chaotic behavior of the electronic motion.
It affects also the magnitude of the enhancement factor. 
We discuss also numerical experiments using polarized targets.   
The derived enhancement factors in our simulation are in agreement with those 
extracted within the $R$-matrix approach.    
\end{abstract}
\end{frontmatter}

\section{Introduction}
The relation between the tunneling process and dynamical chaos has been discussed 
with great interests in the field of nonlinear science~\cite{chaos,kb}.
Though tunneling is a completely quantum mechanical phenomenon, it might be influenced 
by classical chaos.  In the sense that chaos induces fluctuations on the classical 
action which essentially determines the tunneling probability. 
We study the phenomenon by examining the screening effects by bound electrons in low 
energy fusion reactions, where the experimental cross sections 
with gas targets show an increasing enhancement with
decreasing bombarding energy with respect to the values obtained by
extrapolating from the data at high energies~\cite{krauss,eknrsl}. 
Many studies attempted to attribute the enhancement of the reaction
rate to the screening effects by bound target electrons~\cite{bfmmq}. In this context one often 
estimates the screening potential($U_e$) as a constant decrease of the barrier height in the 
tunneling region through a fit to the data. A puzzle has been that the screening 
potential obtained by this procedure exceeds the value of the adiabatic limit, 
which is given by the difference of the binding energies of the united atoms and 
of the target atom and it is theoretically thought to provide the maximum screening 
potential~\cite{rolfs95}. The experiments have been performed using deuterated metallic 
targets, as well, and it is reported that one observes systematically large screening 
potentials for various kinds of metals~\cite{yr}.

The difficulty lies in the determination of the absolute value of the bare cross section.  
The extrapolated bare cross section, i.e., the astrophysical S-factor, might be affected 
by the choice of the high energy region where one assumes the screening potential 
can be neglected~\cite{junker}.
Over these years, several ways to determine the bare cross sections have been proposed.
There are theoretical attempts, using the $R$-matrix theory~\cite{barker}, and  
experimental, using the Trojan Horse Method(THM)~\cite{thm,thm2}. 
However, the THM gives only relative values of the S-factors and one needs to normalize 
to direct methods data obtained at high incident energies.  
Thus the comparison between newly obtained bare cross sections and the cross sections by the 
direct measurements gives a variety of values for the screening potential. 
These values are often smaller than the sudden limit or larger than the adiabatic limit.
Theoretical studies performed 
using the time-dependent Hartree-Fock(TDHF) scheme~\cite{skls,ktab}  
suggest that the screening potential is between the sudden and the adiabatic limits.

One of the aims of this paper is to try to assess the effect of the screening quantitatively. 
Up to now, the dynamical effects of bound electrons have been studied only in some 
limited cases with a few bound electrons(the D+$d$ with atomic target~\cite{skls,ktab} and 
molecular D$_2$ target~\cite{smkls}, the $^3$He+$d$~\cite{skls}) with the TDHF method.  
We investigate here the dynamical effects, including the tunneling region, for other systems 
with many bound electrons; D+D, $^3$He+D, looking the effect of the electron capture of 
projectile, $^6$Li+$d$, $^6$Li+D. 

To simulate the effects of many electrons, we use the constrained 
molecular dynamics (CoMD) model~\cite{kb,pmb,kb-gs}.
At very low energies fluctuations are anticipated to play a substantial role. 
Molecular dynamics contains all possible correlations and fluctuations due to 
the initial conditions(events). 
The prescription using constraints for the Heisenberg uncertainty principle and 
the Pauli exclusion principle is based on the Lagrange multiplier method~\cite{kb-gs}. 
In extending the study to lower incident energies, 
we would like to stress the connection between the motion of bound electrons and chaos. 
In fact, depending on the dynamics, the behavior of the electron(s) is unstable and 
influences the relative motion of the projectile and the target.  
This feature is caused by the nonintegrability of the $N$-body system($N \ge 3$) and it is well known
that the tunneling probability can be modified by the existence of a chaotic environment. 
In this connection we propose a new ``dissipative limit" which is driven by 
electron ejection. We examine also numerical experiments using polarized targets.   

The paper is organized as follows.
In sect.~\ref{sec:form} we introduce the enhancement factor $f_e$ and 
describe the essence of the constrained molecular dynamics approach briefly.
In sect.~\ref{sec:app} we mention limiting 
cases where one can easily estimate the screening potential. We discuss 
particular reactions, the effect of the electron capture by projectile
and the numerical experiments using polarized targets.  
We summarize the paper in sect.~\ref{sec:sum}.

\section{Formalism}
\label{sec:form}
We denote the reaction cross section at incident energy in the center of mass $E$ 
by $\sigma(E)$ and the cross section obtained in absence of electrons by $\sigma_0(E)$.
The enhancement factor $f_e$ is defined as 
\begin{equation}
  \label{eq:fenh}
  f_e\equiv\frac{\sigma(E)}{\sigma_0(E)}.
\end{equation}
If the effect of the electrons is well represented by the constant shift 
$U_e$ of the potential barrier, $\sigma(E)$ is replaced by $\sigma_0(E+U_e)$ 
following~\cite{skls,alr}, 
\begin{equation}
  \label{eq:fenh3}
  f_e\sim\exp\left[\pi\eta(E)\frac{U_e}{E}\right],
\end{equation}
where $\eta(E)$ is the Sommerfeld parameter~\cite{clayton}.
We calculate the enhancement factor through the CoMD simulation.

\subsection{Constrained Molecular Dynamics}
We describe the essence of the CoMD briefly. 
The total Hamiltonian is written down as 
\begin{equation}
H({\bf r};{\bf p})=\sum_i^N ({ \mathcal E}_i+U({\bf r}_i)-m_ic^2),  
\end{equation}
where we use relativistic kinematics: ${ \mathcal E}_i=\sqrt{{\bf p}_i^2c^2+m_i^2c^4},$
$m_i$ and 
\begin{equation}
  \label{eq:coulomb}
  U({\bf r}_i)=\sum^N_{j(\neq i)=0} \frac{q_j q_i }{|{\bf r}_i-{\bf r}_j|}  
\end{equation}
are the energy, the mass and the potential of the $i$-th particle, i.e., 
an electron or a nucleus, respectively. Here, $q_{i}$ is the charge of the particle $i$. 
%and $q_j$ is the charge of electron(for $1\le j\le N$) or nucleus(for $j=0$).  
In the classical molecular dynamics(CMD) one solves the Hamilton equations, i.e.,: 
\begin{eqnarray}
%\begin{align}
  \label{eq:rt}  
%  \label{eq:pt} 
  \frac{d {\bf r}_i}{dt}= \frac{{\bf p}_i c^2}{{\mathcal E}_i};\hspace{0.8cm}
  \frac{d {\bf p}_i}{dt}= -\nabla_{{\bf r}} U({\bf r}_i). 
\end{eqnarray}
%\end{align}
The initial configurations of the bound electrons are prepared using the CoMD, 
as well~\cite{kb-gs}.
To take the feature of the Pauli blocking into account,  
we use the Lagrange multiplier method for constraints.
Our constraints which correspond to the Pauli blocking is 
$\bar f_i\le 1$ in terms of the occupation probability and 
can be directly related to the distance of two particles, i.e.,
$r_{ij} p_{ij}$, in the phase space.
Here $r_{ij}=|{\bf r}_i-{\bf r}_j|$ and 
$p_{ij}=|{\bf p}_i-{\bf p}_j|$. 
The relation $\bar f_i \le 1$ is fulfilled, if $r_{ij} p_{ij}\ge
\xi_P\hbar\delta_{s_i,s_j}$, where $\xi_P=2\pi(3/4\pi)^{2/3}2^{1/3}$, $i,j$ refer 
only to electrons and $s_i,s_j(=\pm 1/2)$ are their spin projection. We can easily 
extend the approach to the Heisenberg principle where the constraint is expressed as
${r}_{ij} {p}_{ij}\ge\xi_H\hbar$, and  $\xi_H=1$, $i$ and $j$ refer to the electrons 
and to the nuclei. Using these constraints, the Lagrangian of the system can be written as 
\begin{equation}
%\begin{eqnarray}
%\begin{align}
%%   {\mathcal L}&=\sum_i \frac{{\bf p}^2_i c^2}{{\mathcal E}_i} 
%%   - U({\bf r}_{i}) \nonumber 
  {\mathcal L}=\sum_i^N {\bf p}_i \cdot \dot{\bf r}_i -H({\bf r}; {\bf p})
  + \sum_{i,j(i)}\lambda^H_i \left( \frac{{r}_{ij} {p}_{ij}}{\xi_H \hbar}-1 \right) 
  + \sum_{i,j(i)}\lambda^P_i \left( \frac{{r}_{ij} {p}_{ij} \delta_{s_i,s_j}}{\xi_P \hbar}-1 \right),  \label{eq:lagc}
\end{equation}
%\end{eqnarray}
%\end{align}
where $\lambda^P_i$ and $\lambda^H_i$ are Lagrange multipliers for Pauli and Heisenberg 
principles respectively. The variational calculus leads to:
\begin{eqnarray}
%\begin{align}
  \label{eq:rt2}  
  \frac{d{\bf r}_i}{dt} &=& \frac{{\bf p}_i c^2}{{\mathcal E}_i} 
  + \frac{1}{\hbar}\sum_{j(i)}\left(\frac{\lambda_i^H}{\xi_H}
    +\frac{\lambda_i^P}{\xi_P}\delta_{s_i,s_j}\right) 
  {r}_{ij}\frac{\partial {p}_{ij}}{\partial {\bf p}_i }, \\
  \label{eq:pt2} 
  \frac{d {\bf p}_i}{dt}&=& -\nabla_{{\bf r}} U({\bf r}_i)
  - \frac{1}{\hbar}\sum_{j(i)}\left(\frac{\lambda_i^H}{\xi_H}
    +\frac{\lambda_i^P}{\xi_P}\delta_{s_i,s_j}\right) 
  {p}_{ij}\frac{\partial {r}_{ij}}{\partial {\bf r}_i }. 
\end{eqnarray}
%\end{align}
where $\lambda^P_i$ and $\lambda^H_i$  are zero, if the two particles $i$ and $j$
are far separated; ${r}_{ij} {p}_{ij}\ge \xi\hbar$ ($\xi=\xi_P$ for electrons $i$ and $j$, 
which have identical spin projections and $\xi=\xi_H$ for the other pairs of particles). 
In this case the eqs.~(\ref{eq:rt2}) and~(\ref{eq:pt2}) redeem the eqs.~(\ref{eq:rt}).      
%% The multiplier $\lambda^P_i$ or $\lambda^H_i$  has finite positive values, 
%% only if $i$ and $j$ closer than $\xi\hbar$. 
The integration of the eqs.~(\ref{eq:rt2}) and (\ref{eq:pt2}) is performed using 
Hermite integration scheme which is efficient and capable of high precision.
%enables integration with high precision. 
The scheme adopts variable and individual time-steps for each electron and nucleus~\cite{ma}.
In this way we obtain many initial conditions which occupy different points in the phase 
space microscopically.   
The Lagrange multipliers are used to obtain the ground states of atomic nuclei. 
During the collision we set the $\lambda_i\equiv 0$ to conserve energy. If needed, a collision 
term among electrons is introduced to avoid the over occupancy of the phase space~\cite{pmb}. 
For the collisions reported here we found the violation of the Heisenberg and Pauli principles  
negligible during the collision dynamics. 

The importance of the influence of the tunneling region to the electron screening has been 
discussed in~\cite{ktab}.
In order to treat the tunneling process in the CoMD framework, we define the collective 
coordinates ${\bf R}^{coll}$ and the collective momentum ${\bf P}^{coll}$ as
\begin{equation}
  {\bf R}^{coll} \equiv {\bf r}_P-{\bf r}_T;   \hspace*{0.5cm}
  {\bf P}^{coll} \equiv {\bf p}_P-{\bf p}_T, 
\end{equation}
where ${\bf r}_T, {\bf r}_P$ ($ {\bf p}_T, {\bf p}_P$) are the coordinates(momenta) of 
the target and projectile nuclei, respectively. 
When the collective momentum becomes zero, we switch on 
the collective force, which is determined by ${\bf F}_P^{coll} \equiv \dot{\bf P}^{coll}$ 
and ${\bf F}_T^{coll} \equiv -\dot{\bf P}^{coll}$, to enter into imaginary time~\cite{bk}.
We follow the time evolution in the tunneling region using the equations,
\begin{equation}
  \label{eq:rti}  
  \frac{d {\bf r}_{T(P)}^{\Im}}{d\tau}= \frac{{\bf p}_{T(P)}^{\Im}}{{\mathcal E}_{T(P)}}; \hspace*{0.5cm} 
  \frac{d {\bf p}_{T(P)}^{\Im}}{d\tau}= -\nabla_{{\bf r}} U({\bf r}_{T(P)}^{\Im}) -2{\bf F}^{coll}_{T(P)},
\end{equation}
where $\tau$ is used for imaginary time to be distinguished from real time $t$.   
${\bf r}^{\Im}_{T(P)}$ and ${\bf p}^{\Im}_{T(P)}$ are position and momentum of the target
(the projectile) during the tunneling process respectively.   
%% Adding the collective force corresponds to inverting the potential barrier
%% which becomes attractive in the imaginary times.    
%The idea comes from that 
In fact in the tunneling region the path which gives an important contribution 
to the action of Feynman path integral lies in the imaginary time region. The path coincides 
with the classical path in the potential valley which is given by turning the barrier upside 
down~\cite{negele}.

The penetrability of the barrier is given by~\cite{bk} 
\begin{equation}
  \label{eq:penet}
  \Pi(E)=\left(1+\exp\left(2{\mathcal A}(E)/\hbar\right)\right)^{-1},
\end{equation}
where the action integral ${\mathcal A}(E)$ is 
\begin{equation}
  {\mathcal A}(E)=\int_{r_b}^{r_a}{\bf P}^{coll}~d{\bf R}^{coll},  
\end{equation}
$r_a$ and $r_b$ are the classical turning points. The internal classical turning point 
$r_b$ is determined using the sum of the radii of the target and projectile nuclei.
Similarly from the simulation without electron, we obtain the penetrability of the bare 
Coulomb barrier $\Pi_0(E)$.
Since nuclear reaction occurs with small impact parameters on the atomic scale,
we consider only head on collisions. 
The enhancement factor is thus given by eq.~(\ref{eq:fenh}), 
\begin{equation}
  f_e=\Pi(E)/\Pi_0(E)
\end{equation}
for each event in our simulation. 
Thus we have an ensemble of $f_e$ values at each incident energy. 

\section{Applications to Electron Screening Problem}
\label{sec:app}

We investigate the enhancement factor for the reactions D+$d$, D+D, $^3$He+$d$, $^3$He+D, 
$^6$Li+$d$, $^6$Li+D.
It is well known that in the low incident energy region the projectile often captures 
electrons before it collides with the target nucleus. 
The atomic projectile cases(D+D, $^3$He+D, $^6$Li+D) in contrast with 
the bare ionic beam projectile cases(D+d, $^3$He+d, $^6$Li+d) reveal the effect 
of the extra electrons. 
%% Actually the screening potentials for many reactions in two well known limit are 
%% tabulated~\cite{bfmmq} including the electron capture. 

\subsection{Two well known limits and the Dissipative limit}
As it is often discussed in the literature, 
there are two limiting cases where one can easily estimate the screening potential
$U_e$. One is the adiabatic and the other is the sudden limit~\cite{bfmmq}.
In the sudden limit one assumes that the relative velocity of the two ions is relatively 
faster than that of the electrons, so that the bound electrons remain frozen during the reaction. 
Instead in the adiabatic limit which is associated with the slow relative velocity 
limit, the electrons change their configurations following the relative motion of ions. 
Especially in the case where the electrons occupy the ground state of the system 
at the beginning of the reaction, the electrons tend to continue occupying the ground 
state throughout the reaction process.  
Effectively the electrons move close to the target nucleus during the collision. 
In such a way the electrons continue to shield the coulomb repulsive field between 
the target and the projectile. 
In this limit the screening potential is given by the difference between 
the binding energies of the target~($B.E.^{(T)}$) and the united atom~($B.E.^{(UA)}$)
which is formed in the last stage of the reaction,
\begin{equation}
  U_e^{(AD)}=B.E.^{(T)}-B.E.^{(UA)}.
\end{equation}
We propose another case which is obtained as a result of the bound electron 
emission. It gives usually a negative effect to the fusion enhancement, because the electrons 
take the kinetic energy away from the ions relative motion.   
We name the new limit as "dissipative limit"(DL).
The screening potential in the dissipative limit is given by 
\begin{equation}
  U_e^{(DL)}=B.E.^{(T)},
\end{equation}
where we assume that the ejected electron has zero kinetic energy.   
If the electron takes finite kinetic energy away, the screening potential  
can be smaller than the dissipative limit.
Note that $U_e^{(DL)}$ has a negative value. 
Such a phenomenon seldom occur in the framework of the mean field 
approach such as TDHF because of the usually assumed spherical (or cylindrical) symmetry of the system. 
We would like to stress here that the DL could be used to perform low energy collisions 
using relatively higher energy beams, similar in some sense to the THM.  
In fact we propose to perform experiments where the fusion residues are detected 
in coincidence with energetic electrons.
Thus increasing the value of the ejected kinetic energy of the electrons is 
as if decreasing the beam energy.

%\subsection{Influence of the tunneling region}

\subsection{Reactions}
\subsubsection{D+$d$ and D+D reactions}
\begin{figure}[htbp]
  \centering
  \includegraphics[width=8.3cm,clip]{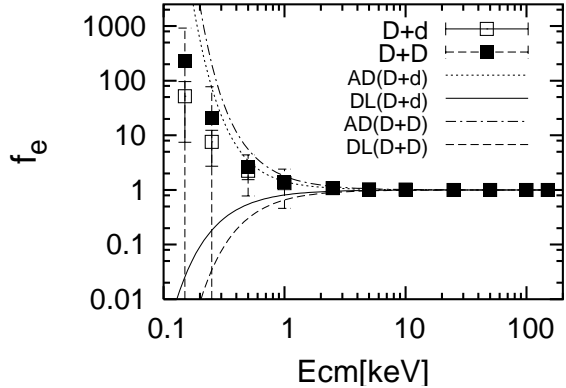}% Here is how to import EPS art
  \caption{Enhancement factor as a function of incident center-of-mass energy 
    for the reactions D+$d$ and D+D. Error-bars represent the variances obtained from the events generated for each beam energy.}
  \label{fig:ef}
\end{figure}

Fig.~\ref{fig:ef}
%, the upper panel
shows the incident energy dependence of the enhancement factor for the reactions 
D+$d$ and D+D, where the systems involve 1 and 2 electrons respectively.
The open and closed squares show the average enhancement factors $\bar{f_e}$ 
over events for the reactions D+$d$ and D+D, respectively.   
The variances $\Sigma=\sqrt{\bar{f_e^2}- (\bar{f_e})^2}$ 
%The variances $\Sigma=\left(\bar{f_e^2}- (\bar{f_e})^2 \right)^{1/2}$ 
are shown with error bars. 
The dotted and dash-dotted curves show the enhancement factors
in the adiabatic limit $f_e^{(AD)}$ for an atomic deuterium target
and it is obtained by assuming equally weighted linear combination of 
the lowest-energy gerade and ungerade wave function for the electron, 
reflecting the symmetry in the D+$d$, i.e., 
\begin{equation}
  \label{eq:fe1}
  f_e^{(AD)}=\frac{1}{2}\left(
    e^{\pi\eta(E)\frac{U_e^{(g)}}{E}}+e^{\pi\eta(E)\frac{U_e^{(u)}}{E}}\right),
\end{equation}
where $U_e^{(g)}=$ 40.7 eV and $U_e^{(u)}=$ 0.0 eV~\cite{ktab,skls} for D+$d$ case. 
If we take into account the electron capture of the projectile, i.e.,
in the case of D+D, the enhancement factor in the adiabatic limit is 
\begin{equation}
  \label{eq:fe2}
  f_e^{(AD)}= \frac{1}{4}
  e^{\pi\eta(E)\frac{U_e^{(g.s.)}}{E}}+\frac{3}{4}e^{\pi\eta(E)\frac{U_e^{(1es)}}{E}},
\end{equation}
where $U_e^{(g.s.)}=$ 51.7 eV and $U_e^{(1es)}=$ 31.9 eV~\cite{kt}. 
The solid curve and dashed curve show the enhancement factors
in the dissipative limit $f_e^{(DL)}$ for the reactions D+$d$ and D+D respectively.
Notice how the calculated enhancement factor with their variances nicely 
ends up between the adiabatic and the dissipative limits.
We performed also a fit of our data using eq.~(\ref{eq:fenh3}) including 
the very low energy region and obtained $U_e=$ 15.9 $\pm$ 2.0 eV for D+$d$ case and 
$U_e=$21.6 $\pm$ 0.3 eV for D+D. 

In Fig.\ref{fig:ef} we saw the rough property of the enhancement factors as average 
values and their variances. Now we discuss their distributions for the reaction 
D+$d$.
\begin{figure}[htbp]
  \centering
  \includegraphics[width=7.3cm,clip]{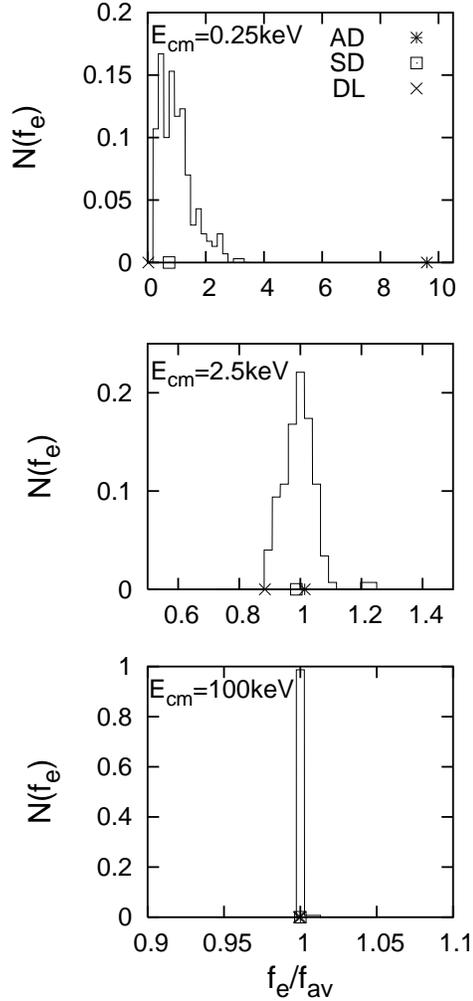}% Here is how to import EPS art
  \caption{Histograms of the normalized enhancement factor for the reaction D+d at various 
    incident energies. In the figure stars, open squares and crosses refer to  the adiabatic, sudden
    and dissipative limits, respectively. }
  \label{fig:hf}
\end{figure}
Fig.~\ref{fig:hf} shows the histograms of the normalized enhancement 
factors($f_e/\bar{f}_e$) at the incident energies $E_{cm}=$0.25keV(top panel), 
2.5keV(middle panel) and 100keV(bottom panel). 
The stars, open squares and crosses specify the adiabatic, sudden and dissipative 
limits at each incident energy, respectively. A remarkable feature in those figures 
is that at the highest incident energy, the enhancement factor for each event in an 
ensemble distribute almost as a $\delta$-function around the average enhancement factor.
As the incident energy goes down, it spreads around the average and the adiabatic limit 
at $E_{cm}=$2.5keV. 
However, at the lowest incident energy $E_{cm}=$0.25keV, where any experiment has not reached   
yet, instead one can see clearly that the distribution of the enhancement factor changes 
and the peak of the distribution is in  between the sudden and the dissipative limits. At a glance it 
appears strange that the enhancement factor goes to the sudden and dissipative limits (which we expected  to be important in the high incident energy region). 
The phenomenon is due to the chaotic behavior of the bound electrons 
in the many-body systems\cite{kb,ss}, which becomes important especially in the extremely 
low energy reactions, where the interaction time is relatively long and the electrons have  
enough time to interact with the projectile ions.

\begin{figure}[htbp]
  \centering
  \includegraphics[width=8.3cm,clip]{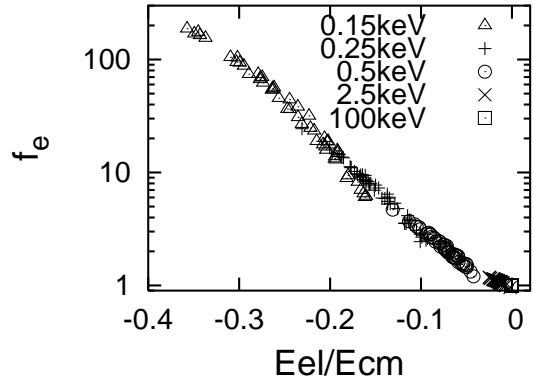}% Here is how to import EPS art
  \caption{The enhancement factor as a function of the binding energies of the electron at 
  the external classical turning point divided by the incident energies for the reaction D+$d$.}
  \label{fig:eef}
\end{figure}
To strengthen the finding of the electrons influence on the enhancement factor, 
in Fig.~\ref{fig:eef} we show the enhancement factors for many events as a function of the
electrons binding energy at the external classical turning point. As one can see, there is a clear 
correlation.
In fact the larger the electrons binding energy, the higher the enhancement factor is.
Thus in order to control the fusion probability we need to have some control on the 
electronic motion especially at low energies as we describe next.      

\subsubsection{Polarized targets}
\label{sec:pt}
\begin{figure}
  \centering
  \includegraphics[width=8.3cm,clip]{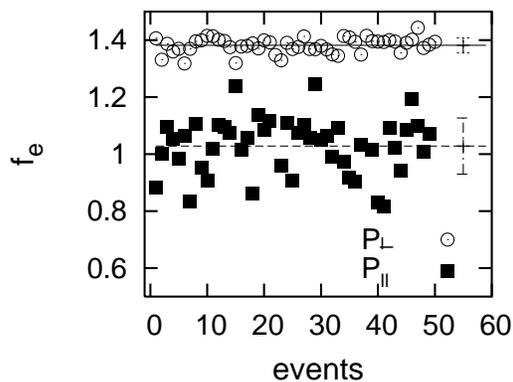}% Here is how to import EPS art
  \caption{The initial configuration dependence of the enhancement factor at the 
    incident energy $E_{cm}=$1.25keV. The enhancement factors for P$_{\perp}$
    targets(open circles) and for P$_{\parallel}$ targets(closed squares)
    are shown.}
  \label{fig:FoEv1}
\end{figure}

We prepared numerically ensembles of target atoms where the electrons motion is 
polarized perpendicular P$_{\perp}$ or parallel P$_{\parallel}$ to the beam axis.   
Fig.~\ref{fig:FoEv1} shows the initial configuration dependence of the enhancement factor
at the incident energy $E_{cm}=$1.25keV.         
The open circles are the enhancement factors for P$_{\perp}$ targets and 
the closed squares are the ones for P$_{\parallel}$ targets.   
It is clear that the P$_{\perp}$ targets give always large enhancement factors 
and a small variance, as it is shown with the error bars in the figure. 
Instead the P$_{\parallel}$ targets give relatively small enhancement factors and large variances. 
As we mentioned in the reference~\cite{kb} or it is discussed in the reference~\cite{ss} as 
an example of the coplanar collisions, this 
is the case where the oscillational force mostly affects the relative motion between the two nuclei.
Because of the non-integrability  of the system under such a force, the motion of 
the electron becomes unstable and it is often ejected to the continuum state. Notice that it 
corresponds to the dissipative limit which we defined in this paper.  
The large variances of the P$_{\parallel}$ targets originates from the fact that the chaotic 
behavior of the system affects the determination of the enhancement factor. 
A remarkable feature is that with the P$_{\parallel}$ targets the enhancement factor often becomes 
less than 1. It means that in this case the bound electron gives an hindrance to 
the tunneling probability. 

\begin{figure}
  \centering
  \includegraphics[width=8.3cm,clip]{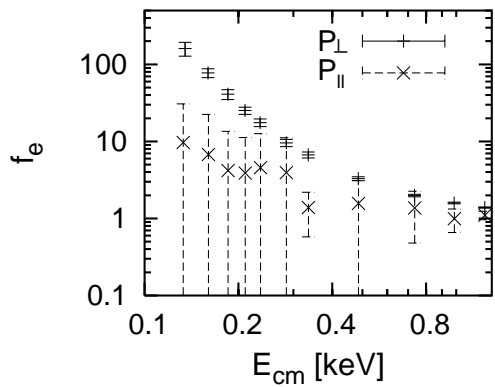}% Here is how to import EPS art
  \caption{same as Fig.~\ref{fig:ef} but for P$_{\perp}$ and P$_{\parallel}$
  targets.}
  \label{fig:EFpol}
\end{figure}

In Fig.~\ref{fig:EFpol} we show the incident energy dependence of the average 
enhancement factor for the P$_{\perp}$ and P$_{\parallel}$ targets with 
pluses and crosses, particularly in the low energy region.  
The enhancement factors from the P$_{\perp}$ targets are always larger 
than that from the P$_{\parallel}$ targets. In contrast to the average enhancement 
from the P$_{\perp}$ targets, which increases monotonically as the incident 
energy becomes smaller, the average enhancement from the P$_{\parallel}$ targets 
fluctuates. It has also large variances at low energies. 

\begin{figure}
  \centering
  \includegraphics[width=7.3cm,clip]{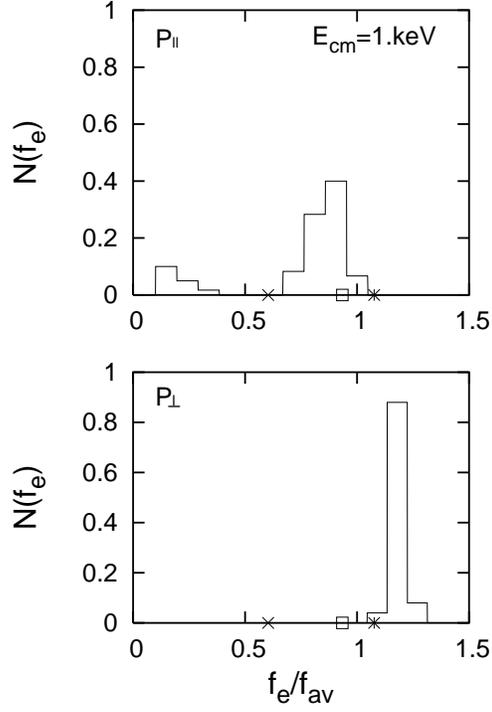}% Here is how to import EPS art
  \caption{Histograms of the normalized enhancement factors for the reaction D+$d$. 
    Incident energy $E_{cm}=$1.keV, polarized parallel target to the beam 
    axis(top panel) and polarized perpendicular target(bottom panel).
    Again stars, open squares and crosses indicate the adiabatic, sudden
    and dissipative limits, respectively.} 
%   and with unpolarized target(bottom panel).}
  \label{fig:N2}
\end{figure}

\begin{figure}
  \centering
  \includegraphics[width=7.3cm,clip]{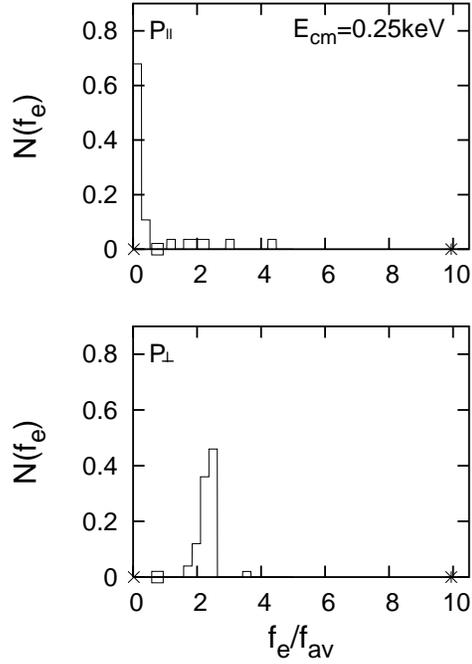}% Here is how to import EPS art
  \caption{same with Fig.~\ref{fig:N2} but at the incident energy $E_{cm}=$0.25keV.}
  \label{fig:N3}
\end{figure}

In figures~\ref{fig:N2} and~\ref{fig:N3} we show the histograms of the enhancement 
factor for P$_{\parallel}$(top panel), P$_{\perp}$(bottom panel) 
at two relatively low incident energies($E_{cm}=$ 1keV and 0.25keV).
In the abscissa the enhancement factor is normalized with its average value over 
the ensemble of unpolarized targets.      

One notices that at the incident energy $E_{cm}=$ 1keV (Fig.~\ref{fig:N2}) the enhancement 
factor for each event distributes around and close to its average value, which is slightly 
less than the adiabatic limit, as a comprehensive feature. Looking carefully at each panel, 
one sees that the P$_{\parallel}$ target  gives smaller enhancement factors than the 
P$_{\perp}$ target, as it is shown in Fig.~\ref{fig:FoEv1}. There is also some  
hindrance in the parallel target case. 

The change from Fig.~\ref{fig:N2} to Fig.~\ref{fig:N3} is drastic. First,
all the events are 
smaller than the adiabatic limit, as we already saw in 
Fig.~\ref{fig:hf} for the case of the unpolarized targets.
Also using P$_{\perp}$ targets we obtain enhancement factors 
always larger than the average value.

\subsubsection{$^3$He+$d$ and $^3$He+D reactions}

An excess of the screening potential was reported for the reactions $^3$He+$d$ with 
atomic gas $^3$He target, and D$_2$ + $^3$He with deuterium molecular 
gas target, for the first time in ~\cite{krauss}.   
Since then various experiments have been performed for these reactions. The incident 
energy covers from 5 keV to 50 keV for $^3$He+$d$. Though once the problem of the discrepancy between 
experimental data and theoretical prediction seemed to be solved by 
considering the correct energy loss data~\cite{lsbr}, recent measurements using measured 
energy loss data~\cite{aliotta} report larger screening potentials than in the adiabatic limit
for both reactions.      

The electron capture by the projectile plays a minor role in the case of $^3$He+d, since
electrons are more bound in helium targets. However 
in the recent measurement Aliotta et al. was performed using molecular D$_2^+$ and 
D$_3^+$ targets~\cite{aliotta}. Thus we assess the contribution from the 
reaction $^3$He+D, as well.     

\begin{figure}[htbp]
  \centering
  \includegraphics[width=8.3cm,clip]{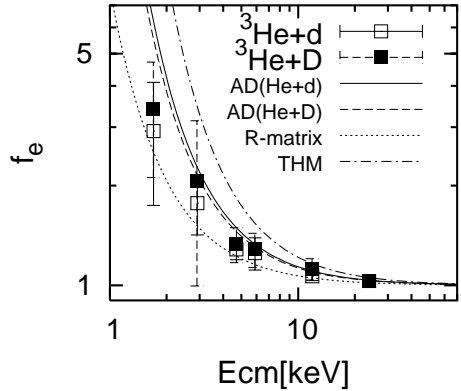}% Here is how to import EPS art
  \caption{Enhancement factor as a function of incident center-of-mass energy 
    for the reactions $^3$He+$d$ and $^3$He+D.}
  \label{fig:ef3Hed}
\end{figure}

The enhancement factor in the adiabatic limit  give 
$U_e$=119 eV for $^3$He+$d$ and $U_e$=110 eV for $^3$He+D, respectively. 
These are shown in the figure~\ref{fig:ef3Hed} with the solid curve 
for $^3$He+$d$ and with the dashed curve for $^3$He+D.
The comparison of these two adiabatic limits implies that the electron capture 
of projectile would give a hindrance compared with the bare deuteron projectile.  
Meanwhile the latest analysis of the experimental data using $R$-matrix two level 
fit~\cite{barker} suggests the screening potential $U_e=$ 60 eV(corresponding 
enhancement factor is shown with dotted curve). 
The comparison between direct measurement and an indirect method, the Trojan Horse 
method, suggests the screening potential $U_e=$ 180$\pm$40 eV ( the corresponding 
enhancement factor is shown with dot-dashed curve)~\cite{thm3}.   
The average enhancement factors $\bar{f_e}$ over events in our simulations using the CoMD 
are shown with  
the open and closed squares for the reactions $^3$He+$d$ and $^3$He+D, respectively.   
The enhancement factors of the both reactions $^3$He+$d$ and $^3$He+D are in agreement 
with the extracted values using the $R$-matrix approach within the variances over 
all the events.  
Notice that our calculated enhancement factors for the two systems display an opposite trend 
as compared to the adiabatic limits.
The average enhancement factor of the reaction $^3$He+D agrees with the estimation of the 
adiabatic limit and the reaction $^3$He+d is below the corresponding adiabatic limit.  
The paradoxical feature comes from the fact that an electron between the two ions is often 
kicked out during the reaction process, i.e., the electron configuration seldom settles down 
the $^5$Li$^+$ ground state in the reaction $^3$He+d. It is known as autoionization in the 
context of the Classical 
Trajectory Monte Carlo method~\cite{gr}. Instead in the case of the $^3$He+D, the  deuterium 
projectile brings its bound electron in a tight bound state around the unified nuclei of 
$^3$He and $d$, practically it ends up with a ground state configuration of the $^5$Li atom.                 
The fits of the obtained enhancement factors suggests the screening potentials 
$U_e=$ 82.4 $\pm$ 1.9 eV for the $^3$He+$d$ and $U_e=$ 102.8 $\pm$ 3.0 eV for the $^3$He+D.
%$U_e=$ 83.2  $\pm$ 1.2 eV for $^3$He+$d$ and $U_e=$ 87.4 $\pm$ 5.3 eV for $^3$He+D.

\subsubsection{$^6$Li+$d$ reaction}

The S-factors for the reactions $^6$Li+$d$, $^6$Li+$p$ and $^7$Li+$p$ were measured over 
the energy range 10 keV $< E_{cm} <$ 500 keV by Engstler,{\it et al.}~\cite{eknrsl}.
They used LiF solid targets and deuteron projectiles as well as deuterium molecular gas 
targets and Li projectiles. 

\begin{figure}[htbp]
  \centering
  \includegraphics[width=8.3cm,clip]{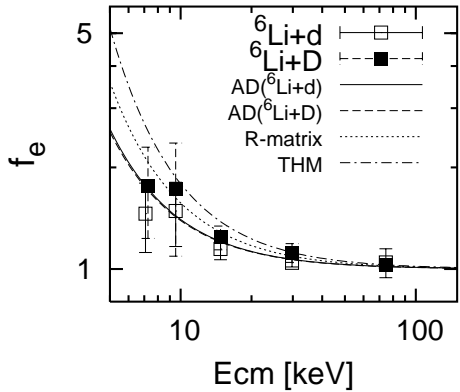}% Here is how to import EPS art
  \caption{same as Fig.~\ref{fig:ef3Hed} but for the reactions $^6$Li+$d$ and $^6$Li+D.}
  \label{fig:ef6Lid}
\end{figure}

In the case of LiF target which is a large band gap insulator, one often approximates 
the electronic structure of the target $^6$Li($^7$Li) state by the $^6$Li$^+$($^7$Li$^+$) 
with only two innermost electrons. Thus for all three reactions one expects 
the screening potential in the adiabatic limit  $U_e^{(AD)}= 371.8-198.2\sim174$ eV. 
Instead if one uses the ground state of the  $^6$Li($^7$Li) atom and of the bare deuteron target 
as the initial state, $U_e^{(AD)}=$186 eV~\cite{bfmmq},  which is given by the solid curve  
in  Fig.~\ref{fig:ef6Lid} .

However one should be aware that the deuteron or hydrogen projectile plausibly moves with 
a bound electron in LiF solid insulator target~\cite{eder}.  
Under such an assumption we could estimate the screening potential 
$U_e^{(AD)}= 389.9-198.2\sim192$ eV.  In the case of molecular D$_2$ or H$_2$ gas targets, 
as well, we should consider  the electron capture by the   lithium projectile.  

The bare S-factors for the same 
reaction have been extracted using an indirect method, the Trojan-Horse Method through 
the reaction $^6$Li($^6$Li,$\alpha\alpha$)$^4$He~\cite{thm2}.
The comparison between direct and the indirect methods gives the screening potential 
$U_e=$ 320$\pm$50 eV. 
The corresponding enhancement factors are shown with the dash-dotted curve. 
The contrast between the direct measurement data and the theoretical estimation for 
the bare S-factor using the $R$-matrix theory gives $U_e$=240 eV. It is shown with dotted line.   
The extracted $U_e$ with the two different methods are larger than the 
adiabatic limit. 

We simulate $^6$Li+$d$ and $^6$Li+D cases only.    
In the figure~\ref{fig:ef6Lid} the open and closed squares show the enhancement 
factor for the reactions $^6$Li+$d$ and $^6$Li+D, respectively. 

Again the average enhancement factors of the reaction $^6$Li+D are larger than those of 
the $^6$Li+$d$. The enhancement factors of the reaction $^6$Li+D are in agreement 
with the extracted values using the $R$-matrix approach within the variances over all the 
events.  
The fit of the obtained average enhancement factors suggests the screening potentials 
%$U_e=$ 176.4 $\pm$ 18.5 eV for $^6$Li+$d$ and $U_e=$ 212.9 $\pm$ 23.9 for $^6$Li+D.
%$U_e=$ 159.0 $\pm$ 14.4 eV for $^6$Li+$d$ and $U_e=$ 211.3 $\pm$ 16.0 for $^6$Li+D.
$U_e=$ 152.0 $\pm$ 9.9 eV for $^6$Li+$d$ and $U_e=$ 214.4$\pm$18.5 for $^6$Li+D.
The screening potential for the reaction $^6$Li+$d$ in our simulation does not exceed the adiabatic 
limit nor extracted values using the $R$-matrix theory and THM, but one for $^6$Li+D verges on the
extracted values using the $R$-matrix approach.

In Table~\ref{tab:ue} we summarize our results of the screening potentials for each reaction 
and compare them with the extracted value using the $R$-matrix theory and THM. 
The screening potentials are obtained by fitting average enhancement factors for each reactions 
and we neglect its variances over all events in this fitting procedure.        
 
\begin{table*}
\caption{Comparison between the screening potential($U_e$) in the adiabatic limit and in our CoMD simulation. The errors in the first column originate from the fitting procedure. \vspace*{.6cm}}
\label{tab:ue} 
\begin{center}
\begin{tabular}{rrrrr}
\hline \hline 
 & & $U_e$(keV) & \\
\cline{2-5}
 & present results & Ref.~\cite{barker}$^{\#}$ & THM & adiabatic limit \\  
% & present results & ref.~\cite{barker}$^\ast$ & THM & adiabatic limit \\  
%\cite{S12}& & \cite{F4}$^\ast$\\ 
\hline 
D+$d$\hspace{0.1cm} &  15.9$\pm$ 2.0 & 8.7, 7.3 & & 22.0$^\ast$ \\[-5pt]  
D+D        &  21.6$\pm$ 0.3 &          & & 37.1$^\ast$ \\ 
$^3$He+$d$\hspace{0.1cm} &  82.4$\pm$ 1.9 & 34, 60, 200 & 180$\pm$40 &119 \\[-5pt]  
$^3$He+D   & 102.8$\pm$ 3.0 &          & & 110 \\ 
%Li$^6$+$d$ & 159.0$\pm$14.4 & 259, 248 & 320$\pm$50 & 186 \\[-5pt]   
$^6$Li+$d$\hspace{0.1cm} & 152.0 $\pm$ 9.9 & 259, 248 & 320$\pm$50 & 186 \\[-5pt]   
%Li$^6$+D   & 212.9$\pm$23.9 &          & & \\ 
$^6$Li+D   & 214.4$\pm$18.5  &          & & \\ 
\hline \hline 
\end{tabular}
\end{center}

\vspace*{.6cm}
\noindent
$^{\#}$ the column of Ref.~\cite{barker} contain not only the results obtained within the 
the $R$-matrix approach but also with a polynomial fitting. \\
$^\ast$ Ref.~\cite{kt}
\end{table*}

\section{Summary}
\label{sec:sum}
We discussed the effect of the screening by the electrons in nuclear reactions 
at the astrophysical energies. 
We performed molecular dynamics simulations with constraints and imaginary time
for the reactions D+$d$, D+D, $^3$He+$d$, $^3$He+D, $^6$Li+$d$, $^6$Li+D. 
For all the reactions it is shown that both the average enhancement factors and their variances 
increase as the incident energy becomes lower. 
Using bare projectiles we obtained the average screening potential smaller than the 
value in the adiabatic limit for all reactions. It is because of the excitation or emission of 
several bound electrons during the reactions. 
The comparison between bare and atomic projectile cases for each reactions revealed that 
the electron capture of the projectile guides to larger enhancements.  
The derived enhancement factors in our simulation are in agreement with those 
extracted within the $R$-matrix approach including the variances over all the events.    
%% The screening potentials which 
%% are obtained by fits of average enhancement factors in the whole incident energy region for 
%% atomic projectile cases are in agreement with that extracted using the $R$-matrix theory. 

We performed numerical experiments using polarized targets for the reaction D+$d$.
Using P$_{\perp}$ targets we obtained relatively large enhancements with small variances, instead 
P$_{\parallel}$ target gives large variances of the enhancement factors and relatively small averaged 
enhancement factors. It is because with the P$_{\parallel}$ targets the force exerted from the electron 
to the relative motion of the nuclei is oscillational, in the direction of the beam axis, and the motion 
of the electron becomes often excited or unstable. It is the case where the chaoticity of the electron 
motion affects the tunneling probability and at the same time the enhancement factor of the cross section.     

{\bf Acknowledgments}

We acknowledge valuable discussions and suggestions with Profs. A.B. Balantekin, T. Motobayashi, 
A. Ono, C. Spitaleri and N. Takigawa. 
One of us(S.K) thanks Prof. T. Maruyama for stimulating discussions and 
for the hospitality at JAERI in Japan.

% === Bishown bliography =====================================================================
%\nocite{*}


\begin{thebibliography}{999}
\bibitem{chaos} W.A.Lin and L.E. Ballentine, {\em Phys. Rev. Lett.\/} {\bf 65} (1990) 2927; 
O. Bohigas, S. Tomosvic, and D. Ullumo, {\em Phys. Rev. Lett.\/} {\bf 65} (1990)5; 
A. Shudo, and K.S.Ikeda, {\em Phys. Rev. Lett.\/} {\bf 74} (1995) 682. 
\bibitem{kb} S. Kimura, and A. Bonasera, {\em Phys. Rev. Lett.\/} {\bf 93} (2004) 262502. 
\bibitem{krauss} A. Krauss, H. W. Becker. H. P. Trautvetter, and C. Rolfs, {\em Nucl. Phys. A\/} {\bf 467} (1987) 273;
\bibitem{eknrsl} S. Engstler {\it et al.}, {\em Phys. Lett. B\/} {\bf 202} (1988) 179.  
\bibitem{bfmmq} L. Bracci, G. Fiorentini, V.S. Melezhik, G. Mezzorani, and P. Quarati, {\em Nucl. Phys. A\/} {\bf 513} (1990) 316.
\bibitem{rolfs95}  C. Rolfs, and E. Somorjai, {\em Nucl. Instrum. Meth. B\/} {\bf 99} (1995) 297.
\bibitem{yr}  H. Yuki {\it et al.}, {\em JETP Lett.\/} {\bf 68} (1998) 823.; 
F. Raiola, {\it et al.}, {\em Phys. Lett. B\/} {\bf 547} (2002) 193.

\bibitem{junker} M. Junker, {\it et al}., {\em Phys. Rev. C\/} {\bf 57} (1998) 2700.
\bibitem{barker} F. C. Barker, {\em Nucl. Phys. A\/} {\bf 707} (2002) 277. 
\bibitem{thm} M. Lattuada, {\it et al}., {\em  Astrophys. J.\/} {\bf 562} (2001) 1076 . 
\bibitem{thm2} A. Musumarra, {\it et al}., {\em Phys. Rev. C\/} {\bf 64} (2001) 068801. 
\bibitem{skls}  T. D. Shoppa, S. E. Koonin, K. Langanke, and R. Seki, {\em Phys. Rev. C\/} {\bf 48} (1993) 837.
\bibitem{ktab} S. Kimura, N. Takigawa, M. Abe, and D.M. Brink, {\em Phys. Rev. C\/} {\bf 67} (2003) 022801(R). 
\bibitem{smkls}  T. D. Shoppa, {\it et al.}, {\em Nucl. Phys. A\/} {\bf 605} (1996) 387.
\bibitem{pmb} M. Papa, T. Maruyama, and A. Bonasera, {\em Phys. Rev. C\/} {\bf 64} (2001) 024612. 
\bibitem{kb-gs} S. Kimura, and A. Bonasera, physics/0409008. 

\bibitem{alr} H. J. Assenbaum and K. Langanke and C. Rolfs, {\em Z. Phys. A\/} {\bf 327} (1987) 461.
\bibitem{clayton} D. D. Clayton, {\it Principles of Stellar Evolution and Nucleosynthesis} 
(University of Chicago Press, 1983) Chap. 4. 
\bibitem{ma} J. Makino, and S. J. Aarseth, {\em PASJ\/} {\bf 44} (1992) 141.
\bibitem{bk} A. Bonasera, and V. N. Kondratyev, {\em Phys. Lett. B\/} {\bf 339} (1994) 207;
T. Maruyama, A. Bonasera, and S. Chiba, {\em Phys. Rev. C\/} {\bf 63} (2001) 057601.   
\bibitem{negele} J.W. Negele, {\em Nucl. Phys. A\/} {\bf 502} (1989) 371. 
\bibitem{kt}  Y. Kato, N. Takigawa, nucl-th/0404075. 
\bibitem{ss} F. Sattin, and L. Salasnich, {\em Phys. Rev. E\/} {\bf 59} (1999) 1246;  
F. Sattin, and L. Salasnich, {\em J. Phys. B\/}, {\bf 29} (1996) L699.    


\bibitem{lsbr} K. Langanke, T.D.Shoppa, C.A.Barnes, and C.Rolfs, {\em Phys. Lett. B\/} {\bf 369} 211(1996).
\bibitem{aliotta} M. Aliotta, {\it et al.}, {\em Nucl.Phys.A\/} {\bf 690}, 790(2001).
\bibitem{thm3} M. La Cognata, {\it et al.}, to be published in Nucl. Phys. A   
\bibitem{gr} T. Geyer and J. M. Rost, {\em J. Phys. B: At. Mol. Opt. Phys.\/} {\bf 36} L107(2003).
and references there in.

\bibitem{eder} K. Eder, et al., {\em Phys. Rev. Lett.\/} {\bf 79} 4112(1997); 
P. Roncin, et al., {\em Phys. Rev. Lett.\/} {\bf 83} 864(1999)

\end{thebibliography}
\end{document}